\documentclass[10pt]{article}
\def\D{\Delta}
\def\d{\delta}
\def\L{\Lambda}

\def\S{\Sigma}
\def\G{\Gamma}

\def\e{\epsilon}
\def\s{\sigma}

\def\i{\iota}
\def\a{\alpha}
\def\b{\beta}

\def\n{\nu}
\def\dim{\textrm{dim}}

\newcommand{\be}{\begin{equation}}
\newcommand{\ee}{\end{equation}}
\newcommand{\bea}{\begin{eqnarray}}
\newcommand{\eea}{\end{eqnarray}}

\begin{document}

\begin{center}
\bf{SPIN FOAM MODELS OF STRING THEORY}\footnote{Talk given at the
Summer School in Modern Mathematical Physics, Zlatibor, 20-31 August 2004. Work supported by the FCT grant POCTI/MAT/45306/2002.}
\end{center}

\bigskip
\bigskip
\begin{center}
A. MIKOVI\'C
\end{center}

\begin{center}\textit{Departamento de Matem\'atica  \\
Universidade Lus\'ofona de Humanidades e Tecnologias\\
Av. do Campo Grande, 376, 1749-024 Lisbon, Portugal\\
E-mail: amikovic@ulusofona.pt}
\end{center}

\bigskip
\bigskip
\begin{quotation}
\small{We review briefly the spin foam formalism for constructing path integrals for the BF and related theories. Then we describe how the path integral for the string theory on a group manifold can be defined as a two-dimensional spin foam state sum. }\end{quotation}

\bigskip
\bigskip
\noindent{\bf{1. Introduction}}

\bigskip
\noindent The spin foam models represent a way to define path integrals for gauge theories as a sum over the irreducible representations (irreps) of the gauge group of the amplitudes for the colored two-complex associated to a triangulation of the spacetime manifold, hence the name spin foam. The prototype spin foam model was the Ponzano-Regge model of 3d Euclidian gravity \cite{pr}, where the partition function is defined as a sum over the $SU(2)$ spins of the product of the $6j$ symbols associated to the tetrahedra of a triangulation. Boulatov then showed that the PR model can be derived by path-integral quantization of the 3d $SU(2)$ BF theory \cite{bu}, and soon after Ooguri showed that the path integral for the 4d $SU(2)$ BF theory can be represented as a sum over the spins of the amplitudes for the colored dual two-complex associated with the manifold triangulation \cite{o}. Independently from these developments, mathematicians Turaev and Viro formulated an invariant for a three-manifold which as an expression was the same as the PR partition function, but instead of the $SU(2)$ group they used the quantum $SU(2)$ group at a root of unity \cite{tv}. Using the quantum group made the divergent PR partition function finite and topologically invariant. This gave a new topological invariant for 3-manifolds. Interestingly, generalizing the TV approach to 4d did not give a new manifold invariant \cite{cyk}.

However, using the spin foam approach for constructing a quantum theory of General Relativity turned out to be more promising \cite{bce,bc,cpr}. Still, this approach is far from being completed \cite{aml}, and more work has to be done. On the other hand, certain lower-dimensional gauge theories can be quite successfully treated via spin foam approach \cite{fk,m}, which
naturally raises the question about the string theory. In the following we describe how the string theory can be reformulated as a spin foam model. 

\bigskip
\bigskip
\noindent{\bf{2. BF theory spin foam quantization}}

\bigskip
\noindent Before discussing the string theory case, let us recall how the BF theory is quantized in the spin foam formalism. The action for this topological theory is given by
\be S= \int_M Tr\,\left( B \wedge F \right) \quad,\ee
where $M$ is a $d$-dimensional manifold, $B$ is a $d-2$ form taking values in the Lie algebra of a Lie group $G$, $F = dA + A\wedge A$ is the curvature two-form for the Lie algebra valued one form $A$ and $Tr$ is the trace. The connection to General Relativity is that the Palatini action can be written as a constrained BF theory for the $SO(3,1)$ group, where the constraint for the $B$ two-form is given by
\be B_{ab}=\epsilon_{abcd} \,e^c \wedge e^d \quad,\label{bcon}\ee
and $e^a$ are the tetrad one-forms. 

The path integral for the BF theory can be defined by discretizing the manifold $M$ with a triangulation, so that one can write
\bea Z &=& \int {\cal D} A\,{\cal D} B\,
\exp\left(i\int_M Tr(B\wedge F)\right)\nonumber\\ &=& \int \prod_l
dA_l \prod_\D dB_\D \exp\left(i\sum_{f} Tr(B_\D F_f
)\right)\quad,\eea where $l$ and $f$ are the edges and the faces
of the dual two-complex $\G$ for the simplical complex (triangulation) $T(M)$, while
$\D$ are the triangles of $T$. The variables $A_l$ and $B_\D$ are defined
as $\int_l A $ and $\int_\D B $ respectively, while $F_f = \int_f F$.

By performing the $B$ integrations
one obtains \be Z= \int \prod_l dA_l \prod_f \delta (F_f) \quad.\ee
This integral can be defined as an integral over the group elements \be Z= \int \prod_l dg_l \prod_f \delta
(g_f) \quad,\ee where $g_l = e^{A_l}$ and $g_f = \prod_{l\in\partial f} g_l$. By using
the well-known identity for the group delta function
\be \delta(g) = \sum_\L \textrm{dim}\,\L \,\chi_\L (g) \quad,\ee
where $\L$'s are the group irreps and
$\chi$'s are the corresponding characters, as well as the integration formula for the representation matrices
\be\int_G dg \, D^{(\L_1)\b_1}_{\a_1}(g)\cdots D^{(\L_n)\b_n}_{\a_n}(g) = \sum_\i C^{\b_1 \cdots\b_n (\i)}_{\L_1 \cdots\L_n}  C_{\a_1 \cdots\a_n}^{\L_1 \cdots\L_n (\i)}\quad,\label{gif}\ee
where $C^{(\i)}$ are the coefficients of the intertwiner $\i$, one obtains
\be Z= \sum_{\L_f,\iota_l} \prod_f \dim\,\L_f \prod_v A_v
(\L_f,\iota_l) \quad.\label{bfss}\ee
The $A_v$ is the vertex amplitude associated to the $d$-simplex dual to
the vertex $v$. This amplitude is given
by the evaluation of the corresponding $d$-simplex spin network, which is in the 3d case given by the $6j$ symbol, while in the 4d case it is given by the $15j$ symbol. Since the sum (\ref{bfss}) is over the irreps labeling the faces of the dual two-simplex, it is called a spin foam state sum \cite{b}.

\bigskip
\bigskip
\noindent{\bf{2. String theory spin foam model}}

\bigskip
\noindent The starting point will be the random triangulation formulation of the bosonic string theory partition function in a flat spacetime \cite{rtr}, which can be written as
\be Z = \sum_T Z(T) = \sum_T \int\prod_{\n} dX_\n \prod_\e e^{-\b(X_{\n(\e)}-X_{\tilde\n(\e)} )^2} \quad,\ee
where $T$ is a triangulation of the two-manifold $\S$, $\n$ are the vertices of $T$, $\e$ are the edges of $T$, $X$ are the string coordinates, $(X_1 - X_2 )^2 = \eta_{ab}(X_1^a - X_2^a ) (X_1^b - X_2^b )$ and $\b$ is a parameter depending on the signature of the flat spacetime metric $\eta_{ab}$. In order to get to a spin foam formulation, we will introduce a discrete Abelian connection variable $A_\e = X_\n - X_{\tilde\n}$. Since this connection is flat, we will have a constraint $F_\D = A_{12} + A_{23} + A_{31} = 0$ for each triangle $\D$ of the triangulation $T$. We can now write
\be Z(T)= \int\prod_{\e} dA_\e \, e^{-\b A_\e^2}\prod_\D \d (F_\D ) \quad.\label{tpf}\ee

Instead of the simplical complex $T$, one can use equivalently the dual complex $\G$. The one-complex for $\G$ is a trivalent graph, and this fact can be used to interpret (\ref{tpf}) as the vacuum Feynman graph for a $Tr\,\phi^3$ matrix field theory with a kinetic term $Tr\,\phi e^{\b \nabla^2}\phi$. This follows from the identification of the momentum $p^a_l$ of the dual edge $l$ as $A^a_\e$, where $\e$ is the corresponding edge of the triangulation $T$. Then
\be Z(T) = Z(\G) =  \int\prod_{l} dp_l \, e^{-\b p_l^2}\prod_v \d (p_v ) \quad,\label{vfd}\ee
where $p_v$ is the sum of the three momenta meeting at the vertex $v$ of $\G$. 

In order to arrive to a spin foam formulation, let us consider now a more general case of a bosonic string propagating on a Lie group manifold $G$, whose action is given by
\be S[X] = \int_\S d^2 \s \sqrt{h} h^{\a\b} Tr\,\left( g^{-1}\partial_\a g g^{-1}\partial_\b g \right) \quad,\ee
where $g(\s)=e^{iX (\s)}$ is a group element, $X$ is the string coordinate and $h_{\a\b}$ is a 2d metric. Then for a dual edge $l$ of a triangulation $T$ we introduce a Lie algebra valued variable $A_l$ given by $A_l =\int_l g^{-1} dg$, so that the expression (\ref{vfd}) is generalized to the non-Abelian case as
\be Z(\G) =  \int\prod_{l} dA_l  e^{-\b Tr\,A_l^2}\prod_v \d (F_v ) \quad,\label{navd}\ee
where $F_v = \int_{\partial\D} g^{-1}dg$.

Let us now rewrite the expresion (\ref{navd}) as the integral over the edge holonomies $g_l = e^{iA_l}$
Let us now rewrite the expression (\ref{navd}) as the integral over the edge holonomies $g_l = e^{iA_l}$
where $g_v$ is the product of the three holonimies meeting at the vertex $v$. We have managed to write the partition function of a string propagating on a group manifold as an integral over the group elements, and this is a starting point for a spin foam formulation.

A gauge-invariant function $f(g)$ satisfies $f(hgh^{-1})=f(g)$, so that it can be expanded as 
\be f(g)=\sum_\L \,\tilde f (\L)\, \chi_\L (g) \quad,\quad 
\tilde f(\L ) = \int_G dg \,\bar\chi_\L (g)\, f(g) \quad.\label{pw} \ee 
This implies that
\be \d (g_v )= \sum_{\L_v} \dim\,\L_v \, \chi_{\L_l}(g_v)\quad,\quad e^{-\b Tr\,A_l^2} = \sum_{\L_l} C(\L_l )\, \chi_{\L_l}(g_l)\quad,\ee
where $C(\L_l)$ will be determined by the integral in (\ref{pw}).
By using the formula (\ref{gif}) for $n=3$ one arrives at 
\be Z(\G) = \sum_{\L_v} \prod_v \dim\,\L_v \left\langle \prod_l A_l (\L_v ,\L_{\tilde v})\right\rangle \quad,\label{nsf}\ee
where the edge amplitude $A_l$ is a matrix
\be A_l (\L ,\tilde\L) = \sum_{\L_l} C(\L_l) C_{\L_l \L \tilde\L}^{\a_l \a \tilde\a}
C^{\L_l \L \tilde\L}_{\a_l \b \tilde\b} \quad.\ee

The evaluation $\langle\,\rangle$ stands for a contraction of the free indices of the $A_l$'s. It is defined by a four-valent graph $\tilde\G$ which is associated to $\G$, in the following way: the number of the vertices of $\tilde\G$ is equal to the number of the edges of $\G$. Then if we associate to each vertex of $\tilde\G$ the amplitude $A_l (\L_v ,\L_{\tilde v})$, the edges of $\tilde\G$ indicate how to make the contractions.

Note that the string action also can have a coupling to a two-form $B_{ab}(X)dX^a \wedge dX^b$, and in the context of the strings propagating on group manifolds, this can be introduced via the Wess-Zumino-Witten term
\be S_{WZW} = \int_M Tr\,\left( g^{-1}dg \wedge g^{-1}dg \wedge g^{-1}dg \right)\quad,\ee
where $M$ is a three-manifold whose boundary is $\S$. This term will induce a term in the discrete string action of the form 
\be \sum_\D \tilde B_{ab}(A) A_{\e}^a A_{\tilde\e}^b \quad.\ee
Since it is not a quadratic function of the edge connections, it can be treated perturbatively, via generating functional technique, e.g. one would need to evaluate the
partition function with sources
\be Z_T (J) = \int\prod_{l} dA_l \, e^{-\b Tr\,( A_l^2 + J_l A_l )}\,\prod_v \d (F_v )
\quad.\ee

This will produce a state sum of the same form as (\ref{nsf}), but now the coefficients $C(\L_l)$ will become functions of the current components $J_l$. The full partition function will be then given perturbatively as
\be \left[e^{\left\langle B\left({\partial\over\partial J}\right) {\partial\over\partial J}{\partial\over\partial J}\right\rangle} Z(J)\right]_{J=0} = Z(0) +\left[ \left\langle B\left({\partial\over\partial J}\right) {\partial\over\partial J}{\partial\over\partial J}\right\rangle Z(J)\right]_{J=0} + \cdots \quad.\ee
 
\bigskip
\bigskip
\noindent{\bf{5. Conclusions}}

\bigskip
\noindent The sum (\ref{nsf}) is a new type of the spin foam sum, and it differs from the usual one by the fact that the vertices of the dual 2-complex are labeled by the group irreps instead of the faces. In the dual picture this means that one labels the triangles of a triangulation, whose weights are given by the dimensions of the corresponding irreps, while the edges have weights as functions of the two triangle irreps who share that particular edge. As a result, the string theory state sum is different from the one coming from the 2d BF theory, and therefore it is not clear how the proposal made in \cite{amstr}, which was based on the 2d BF theory state sum, is related to the standard string theory. 

In order for the sum (\ref{nsf}) to make sense it has to be convergent. We expect this to happen in the Euclidian signature case due to the damping factors $e^{-\b Tr\,A_l^2}$ in the integral (\ref{navd}).  In the Minkowski case a regularization must be made, and the simplest choice is an analytic continuation of the Euclidian result via $\b \to i$. Note that even in the case that the Euclidian state sum is divergent, a regularization can be performed by passing to the quantum group at a root of unity. This can be done because all the simplex weights are the evaluations of certain spin networks, which can be defined in the quantum group case, and the infinite sum over the irreps is replaced by a finite one.

The string partition function for a fixed genus surface will be a sum over the triangulations. It is not known how to do this, so that the next best thing is to consider topological string theories. In that case the partition function is independent of the triangulation, and hence there is no need for a sum

As far as the supersymmetry is concerned, it can be implemented in the spin foam formalism by replacing the Lie group $G$ with a super-group analogue. For example, an
$SO(N)$ group can be replaced by the $OSp(N|2M)$ super-group.

\end{document}